\documentstyle[multicol,eqsecnum,prd,aps,graphicx]{revtex}
\begin{document}
\draft



\title{A Unified Description of Quark and Lepton Mass Matrices\\
 in a Universal Seesaw Model}

\author{\bf Yoshio Koide\thanks{
On leave at CERN, Geneva, Switzerland.}
\thanks{
E-mail address: yoshio.koide@cern.ch; koide@u-shizuoka-ken.ac.jp}
and Hideo Fusaoka\thanks{E-mail address: fusaoka@aichi-med-u.ac.jp
}$^{(a)}$}
\address{
Department of Physics, University of Shizuoka, 
52-1 Yada, Shizuoka 422-8526, Japan \\
(a) Department of Physics, Aichi Medical University,  
Nagakute, Aichi 480-1195, Japan}

\date{\today}

\maketitle
\begin{abstract}
In the democratic universal seesaw model, the mass matrices 
are given by
$\overline{f}_L m_L F_R + \overline{F}_L m_R f_R + 
\overline{F}_L M_F F_R$ ($f$: quarks and leptons; $F$: hypothetical 
heavy fermions), $m_L$ and $m_R$ are universal for up- and 
down-fermions, and $M_F$ has a structure $({\bf 1}+ b_f X)$ 
($b_f$ is a flavour-dependent parameter, and $X$ is a democratic 
matrix).  The model can successfully explain the quark 
masses and CKM mixing parameters in terms of the charged lepton 
masses by adjusting only one parameter, $b_f$. 
However, so far, the model has not been able to
give the observed bimaximal mixing for the neutrino sector. 
In the present paper, we consider that  $M_F$ in the quark sectors
are still ``fully" democratic, while $M_F$ in the lepton sectors are 
partially democratic. Then, the revised model can reasonably give 
a nearly bimaximal mixing without spoiling the previous 
success in the quark sectors.
\end{abstract}


\begin{multicols}{2}

\section{Introduction}

\subsection{What is the universal seesaw model?}

Stimulated by the recent progress of neutrino experiments,
there has been considerable interest in a unified 
description of the quark and lepton mass matrices.
As one of such unified models, a non-standard model, 
the so-called ``universal seesaw model" (USM) \cite{USM}
is well known.
The model describes not only the neutrino
mass matrix $M_\nu$ but also the quark mass matrices
$M_u$ and $M_d$ and the charged lepton mass matrix $M_e$
by seesaw-type matrices, universally:
the model has hypothetical fermions
$F_i$ ($F=U,D,N,E$; $i=1,2,3$) in addition to the conventional 
quarks and leptons $f_i$ ($f=u, d, \nu, e$; $i=1,2,3$), 
and these fermions are assigned to $f_L = (2,1)$, 
$f_R = (1,2)$, $F_L = (1,1)$ and $F_R = (1,1)$ of 
SU(2)$_L \times$ SU(2)$_R$.
The 6 $\times$ 6 mass matrix that is sandwiched
between the fields ($\overline{f}_L, \overline{F}_L$)
and ($f_R, F_R$) is given by
\begin{equation}
M^{6 \times 6} =
\left( \begin{array}{cc}
0 & m_L\\
m_R & M_F
\end{array} \right) \ ,
\end{equation}
where $m_L$ and $m_R$ are universal for all fermion
sectors ($f=u, d, \nu, e$) and only $M_F$ have
structures dependent on the fermion sectors 
$F=U,D,N,E$.
For $\Lambda_L < \Lambda_R \ll \Lambda_S$, where 
$\Lambda_L = O(m_L)$, $\Lambda_R = O(m_R)$ and
$\Lambda_S = O(M_F)$, the $3\times 3$ mass matrix
$M_f$ for the fermions $f$ is given by the
well-known seesaw expression
\begin{equation}
M_f \simeq - m_L M^{-1}_F m_R \ .
\end{equation}
Thus, the model answers the question why the masses of 
quarks (except for top quark) and charged leptons are
so small with respect to the electroweak scale
$\Lambda_L$ ($\sim$ 10$^2$ GeV).
On the other hand, the top quark mass enhancement is 
understood from the additional condition
det$M_F =0$ for the up-quark sector ($F=U$)
\cite{KFzp,KFptp,Morozumi}.
Since the seesaw mechanism does not work for
the third family fermions, the top quark has
a mass of the order of $m_L\sim \Lambda_L$.

For the neutrino sector, the mass matrix is given as
\begin{equation}
\left( \overline{\nu}_L\ \overline{\nu}_R^c\ \overline{N}_L\ 
\overline{N}_R^c \right)
{ 
\left( \begin{array}{cccc}
0 & 0 & 0 & m_L \\
0 & 0 & m_R^{T} & 0 \\
0 & m_R & M_L & M_N \\
m_L^T & 0 & M_N^T & M_R \\
\end{array} \right)
}
\left( \begin{array}{c}
\nu_L^c \\
\nu_R \\
N_L^c \\
N_R
\end{array} \right) \ ,
\end{equation}
where $\nu_R^c \equiv (\nu_R)^c \equiv 
C \overline{\nu}_R^T$.
Since $O(M_N) \sim O(M_L) \sim O(M_R) \gg O(m_R) \gg O(m_L)$,
we obtain the mass matrix $M_\nu$ for the active neutrinos: 
$\nu_L$
\begin{equation}
M_\nu \simeq - m_L M_R^{-1} m_L^T \ .
\end{equation}
If we take the ratio $O(m_L)/O(m_R)$ suitably small,
we can understand the smallness of the observed neutrino
masses reasonably.

For an embedding of the model into a grand unification scenario,
for example, see Ref.\cite{SO10-YK}, where a possibility of
SO(10)$\times$SO(10) has been discussed.

\subsection{What is the democratic universal seesaw model?}

As an extended version of the USM, the ``democratic" USM
\cite{KFzp,KFptp} is also well known. 
The model has successfully given
the quark masses and the Cabibbo--Kobayashi--Maskawa
(CKM) \cite{CKM} matrix parameters in terms of the
charged lepton masses.
The outline of the model is as follows:

\noindent
(i) The mass matrices $m_L$ and $m_R$ have the same structure,
except for their phase factors
\begin{equation}
m_L^f =m_R^f/\kappa =m_0 Z_f \ ,
\end{equation}
where $\kappa$ is a constant with $\kappa \gg 1$ and 
$Z_f$ are given by
\begin{equation}
Z_f =P(\delta_f) Z  \ ,
\end{equation}
\begin{equation}
P(\delta_f) ={\rm diag} ( e^{i\delta_1^f}, e^{i\delta_2^f},
e^{i\delta_3^f} ) \ ,
\end{equation}
\begin{equation}
Z ={\rm diag}\left(z_1,z_2,z_3\right) \ ,
\end{equation}
with $z_1^2 + z_2^2 + z_3^2 = 1$. 

\noindent
(ii) In the basis on which the matrices $m_L^f$ and $m_R^f$ are
diagonal, the mass matrices $M_F$ are given by the form
\begin{equation}
M_F = m_0 \lambda ({\bf 1} + 3b_f X),
\end{equation}
\begin{equation}
{\bf 1} = 
\left( \begin{array}{ccc}
1 & 0 & 0 \\
0 & 1 & 0 \\
0 & 0 & 1 \\
\end{array} \right) \ ,\ \ \ 
X = \frac{1}{3}
\left( \begin{array}{ccc}
1 & 1 & 1 \\
1 & 1 & 1 \\
1 & 1 & 1 \\
\end{array} \right) \ .
\end{equation}

\noindent
(iii) The parameter $b_f$ for the charged lepton sector is
given by $b_e$ = 0, so that in the limit of $\kappa/\lambda
\ll 1$, the parameters $z_i$ are given by 
\begin{equation}
\frac{z_1}{\sqrt{m_e}} = \frac{z_2}{\sqrt{m_\mu}} =
\frac{z_3}{\sqrt{m_\tau}} = \frac{1}{\sqrt{m_e + m_\mu + m_\tau}} \ .
\end{equation}
Then, the up- and down-quark masses are successfully
given \cite{KFzp,KFptp} by the choice of $b_u = -1/3$ 
and $b_d = -e^{i \beta_d}$
($\beta_d = 18^{\circ}$), respectively. 
Here, note that the choice $b_u=-1/3$ gives det$M_U=0$,
so that the case with $b_u = -1/3$ gives $m_t\sim O(m_L)$.
Another motivation for the choice  $b_u=-1/3$ is that 
the model with $b_e=0$ and $b_u=-1/3$ leads to the 
successful relation \cite{Koide-mpl,KFzp} 
${{m_u}/{m_c}}\simeq{({3}/{4})}{({m_e}/{m_{\mu}})}$,
which is almost independent of the value of the seesaw 
suppression factor $\kappa/\lambda$.  
For the choice of $b_u = -1/3$ and $b_d = -e^{i \beta_d}$
($\beta_d = 18^{\circ}$), the CKM matrix parameters are
successfully given \cite{KFzp,KFptp} by taking 
\begin{equation}
\delta_1^u -\delta_1^d=\delta_2^u-\delta_2^d=0 \ ,  \ \ 
\delta_3^u -\delta_3^d \simeq \pi \ .
\end{equation}

A more detailed formulation (including the renormalization
group equation effects) is found in Ref.~\cite{KFevol}.


\subsection{What is the problem?}

It seems that the model is successful as far as the quark
mass phenomenology is concerned, so that the future task is
only to give a more reliable theoretical base to the model.
However, the democratic USM has a serious problem in the 
neutrino phenomenology:
In the previous model, the parameters $z_i$ are fixed by the 
observed charged lepton masses as shown in (1.11), and the 
only adjustable parameter is $b_\nu$ 
defined by (1.9).
For $b_\nu \simeq -1/2$ ($b_\nu \simeq -1$), we can obtain 
the maximal mixing between $\nu_\mu$ and $\nu_\tau$
($\nu_e$ and $\nu_\mu$) \cite{Koide-nu},
while we cannot give the nearly bimaximal mixing, which
is suggested by the observed atmospheric \cite{atm} and
solar \cite{solar,SNO} neutrino data. 

This suggests that the previous model with the universal 
structure of $M_F$ is too tight. Therefore,
in the next section, we assume that for the lepton sectors, 
the democratic matrix $X$ in (1.9) will be changed by 
a ``partially" democratic matrix, which
is given by a rotation $R_X$ from the fully democratic matrix
in the quark sector. Then, we can obtain the observed
nearly-bimaximal mixing. 
However, generally speaking, the success is not so remarkable
because we have three additional parameters in the rotation
matrix $R_X$.
The problem is whether the rotation $R_X$ has a physical
meaning or not.

In Sec.~II, we will investigate a rotation matrix
$R_X$  that leads to the observed nearly bimaximal mixing and 
suggests an interesting relation between quarks and leptons.
In Sec.~III, the numerical results are given and neutrino
phenomenology is discussed. In Sec.~IV, the mysterious 
characteristics of the rotation matrix $R_X$ are discussed.
Finally, Sec.~V is devoted to the conclusions.


\section{S$_2$ symmetry versus S$_3$ symmetry}
\label{sec:2}

\subsection{Basic assumption}

For the quark sectors, the model is essentially unchanged 
from the previous model, i.e. the mass terms are given by
$$
m_0 \sum_{f=u,d} \left[ \overline{f}_L Z  F_R
+ \kappa \overline{F}_L  Z f_R \right.
$$
$$\left.
+\lambda \overline{F}_L P^\dagger(\delta_f)
({\bf 1} +3 b_f X) P(\delta_f) F_R \right] + {\rm h.c}. \ ,
\eqno(2.1)
$$
where we have changed the place of the phase matrix $P$
from  $Z$ to $M_F$, so that $m_L$ and $m_R$ are completely
flavour-independent.
On this basis the mass matrices $m_L$ and $m_R$ are
diagonal, the mass matrix $M_F$ is invariant under the
permutation symmetry S$_3$ except for the phase factors.
As investigated in Refs.\cite{KFzp,KFptp}, in order to give
reasonable values of the CKM matrix parameters, it was
required to choose
$$
P(\delta_u) P^\dagger (\delta_d) =P(\delta_u-\delta_d)
\simeq {\rm diag}(1,1,-1) \ ,
\eqno(2.2)
$$
although the origin of such a phase inversion is still
an open question.
In this paper, we assume
$$
P(\delta_u) ={\rm diag}(1,1,-1) \ , \ \ \ 
P(\delta_d) ={\rm diag}(1,1,1) \ .
\eqno(2.3)
$$

For the lepton sectors, we assume
$$
m_0 \sum_{f=e,\nu} \left[ \, \overline{f}_L Z F_R
+ \kappa \overline{F}_L Z f_R \right.
$$
$$\left.
+\lambda \overline{F}'_L P^\dagger(\delta_f) 
({\bf 1} +3 b_f X) P(\delta_f) F'_R \right] + {\rm h.c.} \ ,
\eqno(2.4)
$$
where, for convenience, we have dropped the Majorana mass 
terms $\overline{N}_L M_L N_L^c + \overline{N}_R^c M_R N_R$
from the expression (2.4), since we always assume that 
the Majorana mass matrices $M_L$ and $M_R$ have the same 
structure as the Dirac mass matrix $M_N= \lambda m_0
P^\dagger (\delta_\nu) ({\bf 1} + 3 b_\nu X)P(\delta_\nu)$.
In (2.4), we have defined
$$
 F'= R_X^T F \ .
\eqno(2.5)
$$
Here, we have tacitly assumed symmetries 
${\rm SU(2)}'_L\times {\rm SU(2)}'_R$ for the heavy
fermions $F_L$ and $F_R$ in addition to the
symmetries ${\rm SU(2)}_L\times {\rm SU(2)}_R$ for $f_L$ and $f_R$,
so that we have required the same rotation $R_X$ for the heavy
leptons $(N_i, E_i)_L$ (and $(N_i, E_i)_R$).
Then, the heavy lepton mass terms in (2.4) can be rewritten as
$$
m_0 \lambda \sum_{f=e,\nu} \overline{F}_L ({\bf 1} +3 b_f {X}_f) 
F_R  + {\rm h.c.} \ ,
\eqno(2.6)
$$
where
$$
X_f =R_X P^\dagger(\delta_f) X P(\delta_f) R_X^T \ .
\eqno(2.7)
$$

We take the phase matrices in the lepton sectors as
$$
P(\delta_\nu) = P(\delta_u)={\rm diag}(1,1,-1) \ ,  
$$
$$
P(\delta_e) = P(\delta_d)={\rm diag}(1,1,1) \ , 
\eqno(2.8)
$$
corresponding to (2.3).
Then, the effective charged lepton and neutrino mass 
matrices are given by
$$
M_e \simeq - m_0 \frac{\kappa}{\lambda}  Z R_X 
({\bf 1}+3 a_e X) R_X^T Z  
$$
$$
\equiv m_0^e Z ({\rm 1} + 3 a_e X_e )Z
\ ,
\eqno(2.9)
$$
$$
M_\nu \simeq - m_0 \frac{1}{\lambda} Z R_X 
P^\dagger (\delta_\nu) ({\bf 1}+3 a_\nu X) 
P(\delta_\nu) R_X^T Z  
$$
$$
\equiv m_0^\nu Z ({\rm 1} + 3 a_\nu X_\nu )Z
\ ,
\eqno(2.10)
$$
where $m_0^e = -m_0 (\kappa/\lambda)$, 
$m_0^\nu = -m_0/\lambda$, $X_e=R_X X R_X^T$ and
$X_\nu = R_X P^\dagger (\delta_\nu) X 
P(\delta_\nu) R_X^T$, and  we have used
$$
({\bf 1} +3 b_f X)^{-1} = {\bf 1} + 3 a_f X 
 \ ,
\eqno(2.11)
$$
$$
a_f = - b_f/(1 +3 b_f) \ .
\eqno(2.12)
$$

The rotation $R_X$ is between the basis in the
quark sectors and that in the lepton sectors.
Our interests are as follows:  What rotation $R_X$ can give
reasonable neutrino masses and mixings?
What relation does it suggest between quarks and leptons?

\subsection{A special form of $R_X$}

In the heavy down-quark mass matrix $M_D$, we have
considered that the matrix $X_d$ is completely
democratic, i.e. $X_d = X$ defined by (1.10).
Hereafter, we define the ``fully" democratic matrix
$X$ defined in (1.10) as $X_3\equiv X$.
The matrix $X_f$ is a rank-1 matrix, which
satisfies the relation $(X_f)^2 = X_f$.
We suppose that the matrices $X_f$ ($f=e, \nu$) 
in the heavy lepton sectors will not be ``fully" 
democratic, but ``partially" democratic. 
The simplest expression of the partially democratic
matrix is
$$
 X_2 \equiv \frac{1}{2}\left(\begin{array}{ccc}
1 & 1 & 0 \\
1 & 1 & 0 \\
0 & 0 & 0 \\
\end{array}\right)
\ .
\eqno(2.13)
$$
We identify $X_e$ as $X_e=X_2$.
The rotation $R_X$, which transforms $X_3$ into $X_2$,
i.e.
$$
R_X X_3 R_X^{T} = X_2 \ ,
\eqno(2.14)
$$
is given by
$$
R_X = R_3(-\frac{\pi}{4})\cdot T \cdot R_3(\theta)
\cdot (-P_3) \cdot A \ ,
\eqno(2.15)
$$
$$
R_3(\theta) = \left( 
\begin{array}{ccc}
\cos\theta & \sin\theta & 0 \\
-\sin\theta & \cos\theta & 0 \\
0 & 0 & 1 \\
\end{array} \right) \ , 
\eqno(2.16)
$$
$$
A = \left( 
\begin{array}{ccc}
\frac{1}{\sqrt{2}} & -\frac{1}{\sqrt{2}} & 0 \\
\frac{1}{\sqrt{6}} & \frac{1}{\sqrt{6}} & -\frac{2}{\sqrt{6}} \\
\frac{1}{\sqrt{3}} & \frac{1}{\sqrt{3}} & \frac{1}{\sqrt{3}} \\
\end{array} \right) \ , 
\eqno(2.17)
$$
$$
T = \left( 
\begin{array}{ccc}
0 & 0 & 1 \\
0 & 1 & 0 \\
1 & 0 & 0 \\
\end{array} \right) \ , \ \ 
P_3 = \left( 
\begin{array}{ccc}
1 & 0 & 0 \\
0 & 1 & 0 \\
0 & 0 & -1 \\
\end{array} \right) \ .
\eqno(2.18)
$$
The matrix $A$ transforms the fully democratic matrix $X_3$
to the diagonal form
$$
A X_3 A^T = \left( 
\begin{array}{ccc}
0 & 0 & 0 \\
0 & 0 & 0 \\
0 & 0 & 1 \\
\end{array} \right) \equiv Z_3 \ . 
\eqno(2.19)
$$
The matrix $Z_3$ is invariant under the rotation $R_3(\theta)$
with an arbitrary $\theta$.
The transformation $T$ has been introduced in order to transform
$Z_3$ to $Z_1 \equiv {\rm diag}(1,0,0)$.
Finally, the rotation $R_3(-\pi/4)$ transforms $Z_1$ to $X_2$.
In the definition of $R_X$, (2.15), we have inserted the
matrix $-P_3$ on the left-hand side of the matrix $A$.
The matrix $-P_3$ does not have any effect on the matrix $Z_3$.
In the numerical study in the next section, we are interested in
the case where $(R_X)_{13}$ takes a small positive value,
so that the matrix $-P_3$ has been introduced to make the
the numerical search easier.

For further convenience, we express the rotation $R_3(\theta)$
by a new angle parameter $\varepsilon = \theta -\pi/4$.
Then, the explicit form of $R_X$ is given by
$$
R_X = \left(
\begin{array}{ccc}
x_3 & x_2 & x_1 \\
\sqrt{\frac{2}{3}}-x_3 & \sqrt{\frac{2}{3}}-x_2 & \sqrt{\frac{2}{3}}-x_1 \\
\sqrt{\frac{2}{3}}(x_1-x_2) & \sqrt{\frac{2}{3}}(x_3-x_1) &
\sqrt{\frac{2}{3}}(x_2-x_3) \\
\end{array}\right)
\ , 
\eqno(2.20)
$$
where $x_i$ are given by
$$
x_1 = \frac{1}{\sqrt{6}} - \frac{c-s}{\sqrt{6}} \ ,
$$
$$
x_2 = \frac{1}{\sqrt{6}} + \frac{c-s}{2\sqrt{6}} - 
\frac{c+s}{2\sqrt{2}} \ ,
\eqno(2.21)
$$
$$
x_3 = \frac{1}{\sqrt{6}} + \frac{c-s}{2\sqrt{6}} + 
\frac{c+s}{2\sqrt{2}} \ ,
$$
($s=\sin\varepsilon$ and $c= \cos\varepsilon$) and
they satisfy the relations
$$
x_1^2+ x_2^2 + x_3^2 = 1 \ ,
\eqno(2.22)
$$
$$
x_1 + x_2 + x_3 = \sqrt{\frac{3}{2}} \ .
\eqno(2.23)
$$


Since we have assumed the inversion $P(\delta_u)$, (2.3),
the heavy up-quark mass matrix $M_U$ (therefore, the matrix
$ P^\dagger(\delta_u) X_3 P(\delta_u)$)
is not invariant under the permutation symmetry S$_3$,
although it is still invariant under the permutation
symmetry S$_2$ for the fields $u_1$ and $u_2$,
because of the form
$$
X_u =  P^\dagger(\delta_u) X_3 P(\delta_u)
= \frac{1}{3} \left(
\begin{array}{ccc}
1 & 1 & -1 \\
1 & 1 & -1 \\
-1 & -1 & 1 
\end{array} \right) \equiv X'_3 \ .
\eqno(2.24)
$$
Since the matrix $X'_3$ is not invariant under the
permutation symmetry S$_3$,
the  neutral heavy lepton mass matrix $M_N$ has a somewhat 
complicated form:
the rank-1 matrix $X_\nu$ is generally given by
$$
X_{\nu} = \left(\begin{array}{ccc}
y_1^2 & y_1y_2 & y_1y_3 \\
y_1y_2 & y_2^2 & y_2y_3 \\
y_1y_3 & y_2y_3 & y_3^2 \\
\end{array}\right) 
\ ,
\eqno(2.25)
$$
where $y_i$ satisfy the normalization $y_1^2+y_2^2+y_3^2=1$.
By comparing the result $R_X X'_3 R_X^T$ from (2.20) with
the expression (2.25), we find
$$
y_1 = \frac{1}{3\sqrt{2}} + \frac{\sqrt{2}}{3} (c-s) \ ,
$$
$$
y_2 = \frac{1}{3\sqrt{2}} - \frac{\sqrt{2}}{3} (c-s) \ ,
\eqno(2.26)
$$
$$
y_3 =  \frac{2}{3} (c+s) \ .
$$

In the next section, we will investigate the
neutrino mass matrix (2.10) numerically.
The expression (2.25) is not always S$_2$-invariant.
Therefore, in the next section, we will require 
the matrix $X_\nu$ to have also an S$_2$-invariant form.
Then, the parameter $\varepsilon$ is fixed, so that
the model can again reduce to a one parameter model
with only $b_\nu$.

\section{Numerical study of the \\
neutrino mass matrix}

In order to find the numerical study of the
neutrino mass matrix (2.10) without spoiling 
the previous success in the quark sectors, 
we evaluate (2.9) in the limit of 
$b_e \rightarrow 0$.
Then, the values of the parameters $z_i$ are still
given by (1.11). Therefore, the numerical success
in the quark sectors \cite{KFzp,KFptp} is unchanged.
The matrix $U_\nu$ by which the mass matrix (2.10)
is diagonalized as
$$
U_\nu^\dagger M_\nu U_\nu^\ast = D_\nu \equiv {\rm diag}
(m_1^\nu, m_2^\nu, m_3^\nu) \ ,
\eqno(3.1)
$$
is the so-called Maki--Nakagawa--Sakata--Pontecorvo
(MNSP) \cite{MNS} matrix. Hereafter, we will simply
call $U_\nu$  the lepton mixing matrix.

The neutrino mass matrix $M_\nu$ has two parameters,
$b_\nu$ and $\varepsilon$.
First, we try to require that the matrix $X_\nu$
be invariant under a permutation symmetry S$_2$.
Although, as suggested from the form $X_e=X_2$ in
 (2.13), the case with $y_1=y_2$ is very interesting,
regrettably it cannot give the observed
nearly-bimaximal mixing for any value of $b_\nu$.
Of the possible cases $y_1=y_2$, $y_2=y_3$ and
$y_3=y_1$, only the case $y_3=y_1$ has a solution
that gives reasonable mixing and mass values.
The case with $y_1=y_3$ fixes the parameters 
$x_i$ and $\varepsilon$
as
$$
y_1=y_3 =0.6900 \ , \ \  y_2=-0.2186 \ ,
\eqno(3.2)
$$
$$
x_1=0.014811 \ , \ \ x_2=0.23904 \ , \ \ 
x_3=0.970890 \ ,
\eqno(3.3)
$$
$$
\varepsilon= 2.043^\circ \ .
\eqno(3.4)
$$ 
As we defined in (2.22) and (2.23), the parameters $x_i$ satisfy the relation
$$
x_1^2 + x_2^2 + x_3^2 =\frac{2}{3} (x_1 +x_2 +x_3)^2 \ .
\eqno(3.5)
$$
On the other hand, it is well known that 
the observed charged lepton
masses satisfy the relation \cite{clmass}
$$
m_e + m_\mu + m_\tau = \frac{2}{3}
\left( \sqrt{m_e} + \sqrt{m_\mu} +\sqrt{m_\tau}\right)^2 \ ,
\eqno(3.6)
$$
i.e.
$$
z_1^2 + z_2^2 + z_3^2 =\frac{2}{3} (z_1 +z_2 +z_3)^2 \ .
\eqno(3.7)
$$
In fact, from relation (3.6), the observed charged lepton masses 
$m_e$ and $m_\mu$ predict $m_\tau^{theor} = 1776.97$ MeV, which is
in excellent agreement with the observed value $m_\tau^{obs}=
1776.99^{+0.29}_{-0.26}$ MeV, together with the parameter values of
$z_i$ for $b_e=0$:
$$
z_1 = 0.016473 \ , \ \ \ z_2 = 0.23687 \ , \ \ \ z_3=0.97140 \ ,
\eqno(3.8)
$$
which correspond to 
$$
\varepsilon=2.268^\circ \ .
\eqno(3.9)
$$ 
It should be noted that the values (3.3) [and (3.4)] are very near 
to the values (3.8) [and (3.9)].
We may consider that the parameters $z_i$ are identical with
the $x_i$, which gives $y_3=y_1$ at a unification
scale $\mu=M_X$.

In the numerical search, the value of the parameter $b_\nu$ is
determined as the prediction $R={\Delta m_{21}^2}/
{\Delta m_{32}^2}$ gives the observed value 
\cite{atm,SNO}
$$
R_{obs} \simeq \frac{5.0 \times 10^{-5} {\rm eV^2}}
{2.5 \times 10^{-3} {\rm eV^2}}
= 2.0 \times 10 ^{-2} \ .
\eqno(3.10)
$$
In Table I, we list the numerical results of 
$b_\nu$, $m_i^\nu$, $\Delta m^2_{21}$, $\Delta m^2_{32}$, 
$\sin^2 2\theta_{12}$, $\sin^2 2\theta_{23}$, and 
$|(U_\nu)_{13}|^2$ as Case A.
Here, for simplicity, we have used the values
$4|(U_\nu)_{11}|^2 |(U_\nu)_{12}|^2$ and 
$4|(U_\nu)_{23}|^2 |(U_\nu)_{33}|^2$ as the values of
$\sin^2 2\theta_{12}$ and $\sin^2 2\theta_{23}$,
respectively, because $R\ll 1$.
For reference, in Table I, we also list a case with
$x_i=z_i = \sqrt{m^e_i/(m_e+m_\mu+m_\tau)}$ as Case B.
In this case, the scenario is that the partially democratic 
form of
$X_\nu$ with $y_3=y_1$ is slightly broken at $\mu=m_Z$, 
still  keeping $x_i=z_i$.
{}From the numerical point of view, there is no essential
difference between the two cases.

The predicted value of $\sin^2 2\theta_{12}$ 
[$\tan^2 \theta_{12}$],
$$
\sin^2 2\theta_{12} = 0.80  \ \ \ [\tan^2 \theta_{12}=0.38] \ ,
\eqno(3.11)
$$
is in good agreement with the present best fit value \cite{SNO}
$\tan^2 \theta_{solar} =0.34$ [$\sin^2 2\theta_{solar}=0.76$].
It should be noted that the predicted value (3.11) gives
a suitable deviation from $\sin^2 2\theta_{12}=1.0$, although
the Zee-type model cannot give such a sizeable deviation from
$\sin^2 2\theta_{12}=1.0$ \cite{Koide-zee}.

It is also worth while noting that in Table I the value of 
$b_\nu$ is very near to $b_\nu =-2/3$.
The results $b_e=0$, $b_u=-1/3$, $b_\nu \simeq -2/3$ and
$b_d \simeq -1$ may suggest the existence of some unified rule
for $b_f$.

Finally, we must excuse ourselves for taking the parameter 
$b_e$ as $b_e \rightarrow 0$ in the numerical calculations.
We have assumed that the heavy charged lepton mass matrix
$M_E$ is given by $M_E = \lambda m_0 ({\bf 1} + 3 b_e X_2)$
on the basis of $F$ (not $F'$), i.e. $M_E$ has the partially
democratic form.
However, the choice $b_e=0$ makes this assumption nonsense.
We consider that the value of the parameter $b_e$ is 
$b_e \simeq 0$, but it is not $b_e=0$.
In fact, although the relation (3.6) has given, for the 
observed charged lepton mass values $m_e$ and $m_\mu$,  
the excellent prediction of the tau lepton mass $m_\tau$,
however, for the values \cite{fusaoka} of $m_e$ and $m_\mu$ 
at $\mu= m_Z$ we obtain the predicted value  
$m_\tau(m_Z)=1724.99$ MeV,
which slightly deviates from the observed value 
$m_\tau(m_Z)=1746.69^{+0.30}_{-0.27}$ MeV \cite{fusaoka}.
This deviation can be adjusted by taking a small deviation of
$b_e$ from zero.

\section{Meanings of the rotation $R_X$}

In the previous section, we have found that the values of 
the parameters $x_i$ with the requirement $y_1=y_3$ are 
very close to the values of $z_i$,
which are evaluated from the observed charged
lepton masses.
It should be noted that only for such a case with 
$x_i\simeq z_i$ we obtain a solution of the value
of the parameter $b_\nu$ that gives reasonable masses
and mixings.
In other words, even if we do not require the 
condition $y_1=y_3$, the phenomenological two-parameter
study with $\varepsilon$ and $b_\nu$ can find a reasonable
solution only when $x_i \simeq z_i$.
This suggests that the rotation $R_X$ has a special meaning
not only for the neutrino mass matrix, but also for the charged 
lepton mass parameter matrix $Z$.
We consider that the coincidence $x_i \simeq z_i$ is
not accidental.

The rotation $R_X$ has the following property:
$$
R_X \left(
\begin{array}{c}
x_3 \\
x_2 \\
x_1 
\end{array} \right) = \left(
\begin{array}{c}
1 \\
0 \\
0 
\end{array} \right) \ ,
\eqno(4.1)
$$
in addition to the property (2.14).
Therefore, it means that the parameters $z_i$ can
be obtained from the vector $(1,0,0)$ by 
the following rotation:
$$
\left(
\begin{array}{c}
z_3 \\
z_2 \\
z_1 
\end{array} \right) = (R_X)^T_{x_i=z_i}\left(
\begin{array}{c}
1 \\
0 \\
0 
\end{array} \right) \ .
\eqno(4.2)
$$

If we define a rotation matrix $\widetilde{R}_X$ as
$$
\widetilde{R}_X = T R_X T \ ,
\eqno(4.3)
$$
where $T$ is defined by (2.18), the relations become
more intuitive:
$$
\widetilde{R}_X X_3 \widetilde{R}_X^T = \widetilde{X}_2
\equiv \frac{1}{2} \left(
\begin{array}{ccc}
0 & 0 & 0 \\
0 & 1 & 1 \\
0 & 1 & 1 
\end{array} \right) \ ,
\eqno(4.4)
$$
$$
(\widetilde{R}_X)_{x_i=z_i} \left(
\begin{array}{c}
z_1 \\
z_2 \\
z_3 
\end{array} \right) = \left(
\begin{array}{c}
0 \\
0 \\
1 
\end{array} \right) \ ,
\eqno(4.5)
$$
$$
(\widetilde{R}_X)_{x_i=z_i}\cdot Z \cdot (3X_3)\cdot 
Z\cdot (\widetilde{R}_X)_{x_i=z_i}^T =
\left(
\begin{array}{ccc}
0 & 0 & 0 \\
0 & 0 & 0 \\
0 & 0 & 1 
\end{array} \right) \ .
\eqno(4.6)
$$

However, in order to obtain the same numerical results
as those in the previous section,
we must change the assumption $X_\nu = R_X P_3 X_3 P_3 R_X^T$
to the following assumption
$$
X_\nu = \widetilde{R}_X P_1 X_3 P_1 \widetilde{R}_X^T \ ,
\eqno(4.7)
$$
where
$$
P_1 = {\rm diag}(-1, 1, 1) \ .
\eqno(4.8)
$$
Then, the parameters $y_i$ in the expression (2.26) 
are given by the same relations with the exchange
between $y_1$ and $y_3$.  
Since we require $y_1=y_3$, the numerical results are
exactly identical with those in the previous section.
In the previous scenario we have assumed,
with the rotation $R_X$, that the heavy up fermions 
take the same
phase matrix $P_3 =P(\delta_u) =P(\delta_\nu)$.
In this case with $\widetilde{R}_X$, we must assume
that $P(\delta_u)=P_3$, but $P(\delta_\nu)=P_1$.
Although the scenario with $\widetilde{R}_X$ is
more intuitive, we cannot at present answer the 
question why quarks require the inversion $P_3$
and why leptons require the inversion $P_1$.

In any case, it is essential that the parameter
values $(z_1,z_2,z_3)$ [or $(z_3,z_2,z_1)$] come
from $(0,0,1)$ [or $(1,0,0)$] by the rotation
$\widetilde{R}_X$ [or $R_X$].
Especially, it is noted that the parameters $z_i$
satisfy the relation (2.23) [therefore (3.7)], 
which leads to the charged lepton mass relation (3.6).
Thus, the rotation $R_X$ has special meanings 
not only as a rotation between the heavy quarks 
$(U,D)$ and $(N,E)$, but also as a rotation
that determines the charged lepton mass
parameters $z_i$.

\section{Conclusions}

We have proposed an improved version of the democratic
universal seesaw model in order to extend the
success of the unified description of the quark and 
charged lepton mass matrices to the neutrino mass
matrix.
In the original model, the mass matrices $m_L$ and
$m_R$ were given by a universal structure $Z$, 
independently of the fermion sectors $f=u,d,e,\nu$,
and the hypothetical heavy fermion mass matrices $M_F$
have the same structure, ``a unit matrix plus a
democratic matrix", which includes only one 
flavour-dependent complex parameter $b_f$.
The constraint was too tight, so that the model could
not give the observed nearly-bimaximal neutrino mixing.
In the improved model, the mass matrices $m_L^f$
(also $m_R^f$) are still flavour-independent, while 
the heavy fermion mass matrices have different
structures between quark and lepton sectors, i.e.
in the quark sectors, $M_F$ still have democratic forms,
while in the lepton sector, $M_F$ have only ``partially"
democratic forms.
If we take a special rotation $R_X$, which transforms
the $3 \times 3$ democratic matrix $X_3$ to the 
$2\times 2$ democratic matrix $X_2$ as (2.14) and if
we take the parameters $x_i$ as $x_i \simeq z_i \propto 
\sqrt{m_i^e}$ and $b_\nu \simeq -2/3$, we can obtain 
reasonable values of neutrino masses and mixings.

For the quark and charged lepton sectors, in the original
democratic universal seesaw model \cite{KFzp,KFptp}, 
we have already obtained reasonable values of the masses 
and mixings by taking $b_e=0$, $b_u=-1/3$, and
$b_d\simeq -1$.
Those values of $b_f$ are unchanged in the present revised 
model and, moreover, in order to explain the observed nearly 
bimaximal neutrino mixing, the value $b_\nu\simeq -2/3$ 
is required.
What is meaning of these parameter values
$$
b_e=0 \ , \ \ \ b_u=-1/3 \ , \ \ \ b_\nu \simeq -2/3 \ ,
\ \ \ b_d \simeq -1 \ ?
\eqno(5.1)
$$
This is a future task for us.

We have also numerically searched a rotation matrix
$R(\theta_{12},\theta_{23},\theta_{31})$ that can give
reasonable values for the observed neutrino mixings and
masses, without requiring the constraint (2.14).
We found that the only solution is the rotation
$R_X$ with $x_i\simeq z_i$ (the values $z_i$ are given
by (1.11)) for $b_\nu \simeq -2/3$.
The solution $R_X$ transforms the ``fully" democratic
matrix $X_3$ into the partially democratic matrix $X_2$
and the parameters $x_i$ satisfy the relation (3.5),
which leads to the charged lepton mass formula (3.6).
The rotation $R_X$ with $x_i\simeq z_i$ also transforms
the matrix $X'_3$ (2.24) into a partially democratic
matrix $X_\nu$ (2.25) with $y_1 =y_3$.
These mean that the observed neutrino data require
not a mere numerical solution of 
$R(\theta_{12},\theta_{23},\theta_{31})$, but the
special solution $R_X$ with $x_i=z_i$.
The observed charged lepton masses, which are
proportional to $z_i^2$, are closely related to
the rotation $R_X$ with $x_i=z_i$, for example
as (4.2), (3.7), and so on.
These facts give us a sufficient motivation for
the rotation $R_X$ with $x_i=z_i$ be
taken seriously.
However, at present, the theoretical origin of 
the rotation is not clear.
This is also a future task to us.

\vspace{3mm}
{\bf Acknowledgement}

The authors wish to acknowledge the hospitality of the theory
group at CERN, where this work was completed.


\end{multicols}

\vspace{1cm}
\begin{table}
\begin{center}

\begin{tabular}{|c|c|c|}\hline
      & Case A with $x_i\neq z_i$ & Case B with $x_i=z_i$ \\ \hline
$b_\nu$  &  $ -0.680$    & $ -0.684$     \\ \hline
$a_\nu$  &  $ -0.654$    & $ -0.650$     \\ \hline
$m_1^\nu$ (eV) & $ 2.39 \times 10 ^{3} $ & 
                 $ 2.43 \times 10 ^{3} $ \\
$m_2^\nu$ (eV) & $ 7.46 \times 10 ^{3} $ &
                 $ 7.48 \times 10 ^{3} $ \\
$m_3^\nu$ (eV) & $ 5.06 \times 10 ^{2} $ &
                 $ 5.06 \times 10 ^{2} $ \\ \hline
$\Delta m^2_{21}$ (eV$^2$) & $ 5.00 \times 10 ^{-5} $ &
                             $ 5.01 \times 10 ^{-5} $  \\
$\Delta m^2_{32}$ (eV$^2$) & $ 2.50 \times 10 ^{-3} $
                           & $ 2.50 \times 10 ^{-3} $ \\ 
$\Delta m^2_{21}/\Delta m^2_{32}$ & $ 2.00 \times 10 ^{-2}$ & 
                            $ 2.00 \times 10 ^{-2}$ \\ \hline
$ \sin^2 2\theta_{12}$ & 0.796 & 0.801 \\
$( \tan^2 \theta_{12})$ & (0.377) & (0.383) \\
$ \sin^2 2\theta_{23}$ & 0.978 & 0.979 \\
$|(U_\nu)_{13}|^2$ & $6.65 \times 10 ^{-3}$  &
 $6.68 \times 10 ^{-3}$ \\ \hline
\end{tabular}
\caption{ 
Predictions of the neutrino masses and mixing parameters.
For the predictions $\Delta m^2_{ij}$ and $m_i^\nu$,
we have used the value $\Delta m^2_{32}=2.5\times 10^{-3}$
eV$^2$ from the atmospheric neutrino data \protect\cite{atm}.
In case A, the values $x_i$ are determined from the
requirement $y_1=y_3$, and the values $z_i$ are obtained
from the relation (3.7) and the observed values of $m_e$
and $m_\mu$.  In case B, the values $x_i$ are taken
as $x_i=z_i$, where $z_i$ are obtained as in case A.
}
\end{center}

\label{I}

\end{table}


\end{document}